\documentclass[sn-mathphys,referee]{sn-jnl}

\usepackage{amsmath}
\usepackage{bm}
\usepackage{graphicx}
\usepackage{float}
\usepackage{cancel}
\usepackage{placeins}
\usepackage{caption}
\usepackage{xcolor}
\usepackage{lineno}
\usepackage{ulem}

\jyear{2021}%

\theoremstyle{thmstyleone}%
\theoremstyle{thmstyletwo}%

\theoremstyle{thmstylethree}%

\bibstyle{sn-mathphys}

\begin{document}
\title[Multi-photon electron emission with non-classical light]{Multi-photon electron emission with non-classical light}

\author[1]{Jonas Heimerl}
\equalcont{These authors contributed equally to this work.}
\author[2]{Alexander Mikhaylov}
\equalcont{These authors contributed equally to this work.}
\author[1]{Stefan Meier}
\author[1]{Henrick Höllerer}
\author[3]{Ido Kaminer}
\author[1,2]{Maria Chekhova}
\author[1]{Peter Hommelhoff}

\affil[1]{Department of Physics, Friedrich-Alexander-Universität Erlangen-Nürnberg (FAU), 91058 Erlangen, Germany}
\affil[2]{Max Planck Institute for the Science of Light, 91058 Erlangen, Germany}
\affil[3]{Department of Electrical and Computer Engineering, Russell Berrie Nanotechnology Institute, Technion–Israel Institute of Technology, Haifa 32000, Israel}

\affil[]{e-mail: jonas.heimerl@fau.de, peter.hommelhoff@fau.de}

\date{\today}

\abstract{
Photon number distributions from classical and non-classical light sources have been studied extensively, yet their impact on photoemission processes is largely unexplored. 
In this article, we present measurements of electron number-distributions from metal needle tips  illuminated with ultrashort light pulses of different photon quantum statistics. By varying the photon statistics of the exciting light field between classical (Poissonian) and quantum (super-Poissonian), we demonstrate that the measured electron distributions are changed substantially.
Using single-mode bright squeezed vacuum light, we measure extreme statistics events with up to 65 electrons from one light pulse at a mean of 0.27 electrons per pulse -  the likelihood for such an event equals $10^{-128}$ with Poissonian statistics. Changing the number of modes of the exciting bright squeezed vacuum light, we can tailor the electron-number distribution on demand. Most importantly, our results demonstrate that the photon statistics is imprinted from the driving light to the emitted electrons, opening the door to new sensor devices and to strong-field quantum optics with quantum light.
}

\maketitle

In daily life, numerous processes adhere to Poissonian statistics, yet we often encounter extraordinary occurrences in certain domains like climate or the trade market that defy such statistical norms~\cite{Sura2011, Fernandez2005}. Modelling such events in real-life systems is of enormous importance but often a challenge~\cite{albeverio2006extreme}. Therefore, well controllable model systems in non-linear optics have attracted much attention to study such extreme events. For example, optical "rogue" waves have been created in non-linear fibers~\cite{Solli2007}, in close analogy to oceanology~\cite{Kharif2003}.  The investigated distributions share a common characteristic: while most events exhibit small amplitudes, a substantial likelihood exists for  observing events with extreme amplitudes exceeding the mean value considerably. This likelihood is typically orders of magnitudes smaller than predicted by Poissonian statistics. 

Quantum light generated from spontaneous parametric down-conversion in non-linear crystals is ideally suited to enter the highly non-Poissonian realm. When strongly pumped, the generated spontaneous parametric down-conversion light  becomes so bright that it is visible by eye and can even generate optical harmonics, hence it is called bright squeezed vacuum (BSV)~\cite{mandel1995optical,Manceau2019}. Because of the correlated photon pair generation BSV light shows superbunched photon statistics. For the pulsed case, BSV light is characterized by intense inter-pulse and intra-pulse photon-number fluctuations, which lead to a very broad-tailed photon-number distribution \cite{Manceau2019}.

Extreme events are inherently extremely interesting to study, yet their interaction with matter becomes even more intriguing. In particular, do the extreme counting statistics prevail, and is it even enhanced in non-linear interactions? We here investigate how such highly fluctuating quantum light interacts with matter. With this light-matter interaction we enter a new realm since this light cannot be described as a classical field any longer~\cite{Gorlach2022}. More specifically, we study non-linear multiphoton photoemission from metal needle tips using pulsed BSV light, compared to pulsed coherent light that serves as the reference with Poissonian photon statistics. We demonstrate that the electron number distribution resulting from BSV light exhibits events even more extreme than the driving BSV light itself: We observe extreme electron events exceeding the mean value by a factor of 240. This behavior can be attributed to the interplay of BSV driving light and the non-linearity of the emission process. We show that the photon number statistics of the exciting light are transferred to the emitted electron number distribution. Tailoring the number statistics of electron beams directly in the emission process is particularly intriguing for the field of (electron-based) imaging. We foresee that imprinting {\it sub}-Poissonian photon statistics will enable shot noise-reduced imaging in the future, a highly important feature for interaction-lean imaging of   biological samples~\cite{Henderson1995,Okamoto2012,Kruit2016,Berchera2019}.

\begin{figure}[ht]
	\centering
\includegraphics[width = 0.99\linewidth]{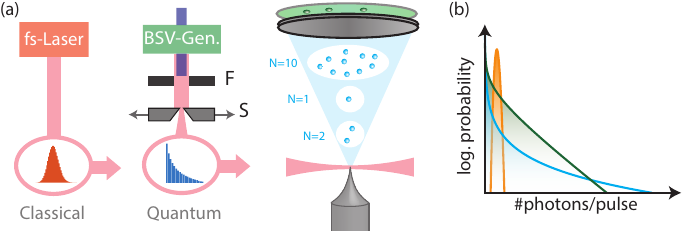}
\caption{(a) Sketch of the optical setup with the femtosecond erbium fiber laser system and bright squeezed vacuum source based on parametric down conversion. To adjust the number of modes in the BSV light, spectral (F) and spatial filtering (S) is required, which we realize with variable bandpass filters and a variable slit (see text for details). Both pulsed beams are focused on the apex of a tungsten needle tip, in separate measurements. The emitted pulses of electrons (blue dots) are accelerated from the tip towards a micro-channel plate detector with a phosphor screen. We obtain the electron number statistics by counting the number of electrons $N$ emitted from each light pulse. (b) Schematic representations of photon-number distributions for the coherent (orange) and the BSV light (green, first-order Gamma distribution)  with the same mean number of photons per pulse. We also show the fourth-order Gamma distribution (blue), which exhibits an even heavier tail (see text for details). }
\label{fig:setup}     
\end{figure}

In the experiment we use two light sources: a pulsed laser beam emitting coherent light and a non-classical BSV light source. The two sources are used alternatively as  depicted in Fig.~\ref{fig:setup}(a). The coherent light source is a femtosecond erbium fiber laser with a central wavelength of 1550\,nm and pulse duration of 170\,fs. The BSV light source is based on parametric down-conversion in a 10\,mm long type-I beta-barium borate (BBO) crystal. The crystal is pumped with 800\,nm laser light with a pulse duration of $\sim$1.6\,ps. The resulting BSV spectrum spans a broad range from 1400\,nm to 1900\,nm with a central wavelength of 1600\,nm. The pulse repetition rate for both sources is set to $\sim$100\,Hz by pulse picking.

The two light sources display vastly different photon number distributions. For coherent light the photon number distribution is Poissonian~\cite{mandel1995optical} [Fig.~\ref{fig:setup}(b)]. In contrast, the light statistics of BSV can be altered: When we gradually change the spatial and spectral filtering of the BSV light, we go from multimode to the (nearly) single mode case. This allows us to vary progressively its photon number statistics. In the case of one spatial and spectral mode, the BSV light's photon statistics obeys a Gamma distribution~\cite{Manceau2019}
\begin{linenomath}
\begin{equation}
    P_{\mathrm{BSV}}(N) = \frac{1}{\sqrt{2\pi \langle N \rangle N}}e^{-\frac{N}{2\langle N \rangle}},
    \label{eq:gamma_1}
\end{equation}
\end{linenomath}
a broad and heavy-tailed distribution with the highest probability of {\it zero} photons per pulse always [Fig.~\ref{fig:setup}(b), green], where $N$ is the number of photons and $\langle N \rangle$ the average of photons per laser pulse. In the multi-mode case, the tail becomes smaller for an increasing number of modes \cite{Allevi2015}.

To adjust the number of spectral modes in the BSV  light beam, we  use a bandpass filter with 50\,nm full width at half maximum (FWHM), with the central wavelength matching the degenerate wavelength of the BSV spectrum at 1600\,nm. We place a slit in the back Fourier plane of the lens that focuses the pump laser light into the crystal [Fig.~\ref{fig:setup}(a)]. This slit allows us to filter the spatial modes. We ensure that the 800\,nm pump light is well suppressed with a dichroic mirror and bandpass filters placed after the BSV generation stage.

Each beam can be sent to our vacuum chamber, where it is focused by an off-axis parabolic mirror (OAP) onto a sharp tungsten needle tip with a typical apex radius of $\sim 10$\,nm. The tip is situated in an ultra-high vacuum chamber with a base pressure of $1\cdot 10^{-9}$\,hPa [Fig.~\ref{fig:setup}(a)]. The beam size for both beams is $\sim 5\,\mathrm{\mu m}$ ($1/e^2$ intensity radius). This rather tight focusing allows us to reach peak intensities of up to $\sim 5\cdot 10^{12}\,\mathrm{W/cm^2}$ for the coherent case (Gaussian pulse) assuming a field enhancement of a factor of 6 at the tip apex~\cite{Thomas2013}. The resulting intensity is sufficient to trigger notable electron emission via a non-linear multiphoton process. For BSV light, the available peak intensity depends on the filtering and the resulting ultrafast fluctuations. For unfiltered BSV light, the intensity can reach up to $3\cdot 10^{12} \,\mathrm{W/cm^2}$, assuming a Gaussian envelope in time. 

The emitted electrons are accelerated from the negatively biased tip ($-200$\,V) towards a Chevron-type micro-channel plate detector (MCP) with a phosphor screen, serving as a spatially resolving single electron detector: Each electron amplified by the MCP generates a bright spot on the phosphor screen, which is imaged with a CCD camera, whose frame acquisition is synchronized to the repetition rate of the excitation light pulses. To retrieve the electron number statistics, we determine the electron number $N_e$ (number of bright spots) for each recorded frame, i.e., each light pulse, via an image recognition code. This way we register up to several hundred of electrons per pulse. Typically, we record around $1\cdot 10^4-4\cdot10^4$ frames for each excitation power, corresponding to a few minutes of measurement time. We use these frames to determine the average number of detected electrons per pulse, $\mu$. The MCP shows a background of $\mu \sim$ 0.002 counts per pulse, independent of the incoming electrons.

\begin{figure*}[ht]
   \centering
     \includegraphics[width = 0.6\linewidth]{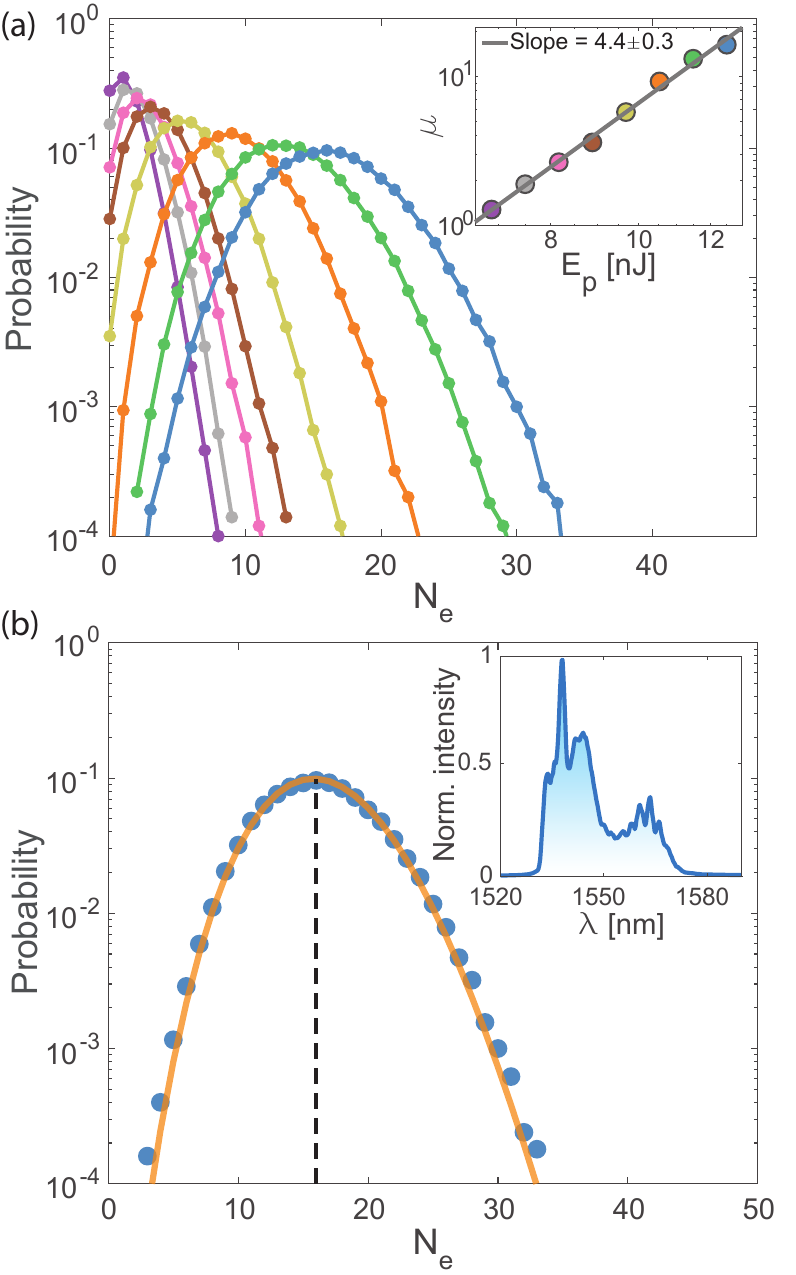}
      \caption{Electron number statistics with coherent excitation light. Panel (a) shows the electron number distributions, namely the detection probability plotted versus the number of detected electrons $N_e$, for various pulse energies. The maxima of the distributions shift towards higher mean values for increasing excitation light energy. The lines connecting the dots serve as a guide to the eye. The inset shows the double-logarithmic plot of the mean number of the registered electrons per pulse $\mu$ versus the pulse energy $E_p$. The color of each data point here matches the color of the electron number distribution in the main graph. The slope of the curve obtained from fitting the inset data (solid grey line) indicates the non-linearity of the photoemission process as $n = 4.4\pm0.3$. (b)  Example electron number distribution for a mean value of $\mu = 16.0$ and $E_p = 13$\,nJ taken from (a). The orange line shows the Poisson distribution calculated for $\mu = 16$. We observe almost perfect agreement between the analytical curve and the experiment, without any free parameter adjustment. Hence, the number statistics remains Poissonian as expected. The inset shows the spectrum of the coherent light source, where the horizontal axis is the wavelength $\lambda$ and the vertical axis is the normalized intensity.}
  \label{fig:Poisson}
\end{figure*}

Fig.~\ref{fig:Poisson} shows the results obtained with coherent light excitation. It shows a histogram of the measured probability distribution of the number of detected electrons versus the number of detected electrons per pulse $N_e$ [Fig.~\ref{fig:Poisson}(a)]. The probability is normalized to the total electron counts. Different curves (colors) correspond to different pulse energies $E_p$, spanning the range of $7\dots 13$\,nJ. We observe that the maximum shifts towards higher average values $\mu$ with increasing pulse energies: from $\mu = 1$ at $E_p = 7$\,nJ to $\mu = 16$ at $E_p=13$\,nJ, as shown in the inset. A fit to this data yields a non-linearity of $n = 4.4\pm0.3$, manifesting the four-photon excitation nature as the dominating contribution in the emission.

In Fig.~\ref{fig:Poisson}(b) we pick the distributions with $\mu = 16.0$ ($E_p$= 13\,nJ) and overlay an analytical Poisson distribution of the same mean (orange)  for comparison. We note that the analytical curve matches the data almost perfectly, without any fitting. This excellent agreement demonstrates experimentally that the multi-photon photoemission shows Poissonian electron number distributions, at well above one electron per pulse. Event-resolved electron number distributions have so far only been shown for single-photon photoemission or in conjunction with electron optics, influencing the number statistics, with well below one electron per laser pulse~\cite{haindl2022coulomb, meier2022few}.

The Poissonian nature of the electron emission implies that each electron is emitted independently, even when there is clearly more than one electron per pulse emitted. In particular, we observe no hint for any blockade mechanism even for large $N_e$ during the emission process caused by either Coulomb interaction or Pauli exclusion, a surprising finding given the nanometer-femtosecond emission dimensions, and the fact that recently Coulomb-induced two-electron energy correlations have been shown to be rather strong~\cite{haindl2022coulomb, meier2022few}. 
    
\begin{figure*}[ht]
\includegraphics[width = 0.8\textwidth]{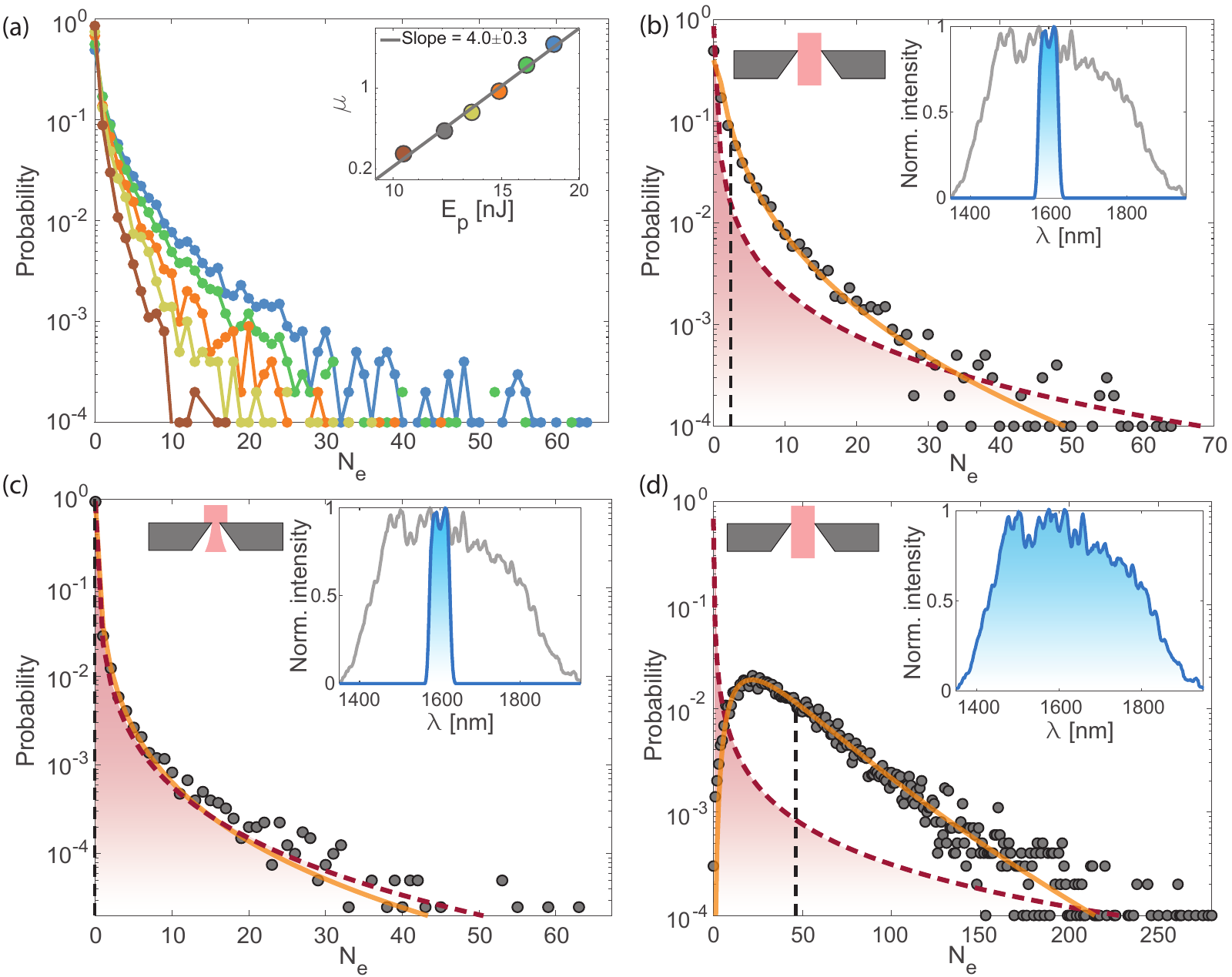}
\caption{Electron number statistics with bright squeezed vacuum light (BSV). (a) Electron number distributions for various BSV pulse energies. Colors match the inset colors. Notably, all curves show the highest probability at $N_e = 0$. The BSV light was filtered with a 50\,nm wide bandpass around the degenerate wavelength of 1600\,nm. The inset shows the non-linear scaling of the mean number $\mu$ of emitted electrons per laser pulse versus the pulse energy $E_p$ on a double logarithmic scale. The fit yields a slope of $n= 4.0.\pm0.3$, similar to the coherent case. (b)-(d) show the effect of different spatial and spectral mode filtering on the electron number distribution, as  indicated by the slit size (upper left insets) and by the spectral filtering (upper right insets). The analytical curves (dashed red) always depict the single-mode case ($m$=1) for a fourth-order non-linearity ($n=4$), i.e., they show a fourth-order Gamma distribution for the measured mean value (black dashed line). (b) Electron number distribution with experimental mean value $\mu = 2.6$ from (a) for the highest pulse energy of $E_p$= 18\,nJ , showing the best agreement for $m=11\pm3$ modes (orange). (c) With spatially and spectrally filtered BSV light together with the theoretical curve for $m=1$ (red) and  $m=2$ (orange). Extreme events of up to 65 electrons per pulse are observed, several orders of magnitude higher than the mean value $\mu = 0.27$, demonstrating the extreme fluctuations of BSV light imprinted on the photo-emitted electrons. (d) Using completely unfiltered BSV light, mean values as high as $\mu = 48$ can be observed. Due to uncompensated group velocity dispersion in the optics beam path the number of modes increases to $m = 57\pm5$ (see text for details). In stark contrast to the single-mode case, the distribution maximum is now clearly shifted away from 0.}
\label{fig:BSV}
\end{figure*}

We now use BSV light to trigger the photoemission. Fig.~\ref{fig:BSV}(a) shows the obtained electron number distributions for various BSV pulse energies $E_p$ in the range of $9 \dots 18$\,nJ, like above. Here, the BSV light is filtered spectrally with a 50\,nm bandwidth filter at the degenerate wavelength, but so far not spatially. We observe that events with higher electron counts $N_e$ (extreme events) become more likely as we increase the pulse energy, leading to broader distributions. Yet and notably, the highest probability remains at $N_e =0$ for all pulse energies. The inset of Fig.~\ref{fig:BSV}(a) shows the non-linear scaling of the mean values $\mu$ versus the excitation energies $E_p$. Fitting of this dependence yields a slope value of $n=4.0\pm0.3$, similar to the coherent case.

Two parameters are crucial to understand the obtained BSV electron number distribution in detail, namely the non-linearity of the emission process $n$ and the number of optical modes $m$ present at the tip apex. To accommodate the non-linearity of an $n$-photon process pumped by a light with a Gamma distribution Eq. (\ref{eq:gamma_1}) of photon numbers, a generalized  $n^{\mathrm{th}}$-order Gamma function is required (see~\cite{Manceau2019} for details):
\begin{linenomath}
\begin{equation}
P_n (N) = \frac{\sqrt[2n]{(2n-1)!!}}{n\sqrt{2\pi}\sqrt[2n]{\langle N \rangle}N^{1-1/2n}}e^{-\frac{1}{2}\sqrt[n]{(2n-1)!!\frac{N}{\langle N \rangle}}}.
\label{eq:gamma_2}
\end{equation}
\end{linenomath}
The overall shape of the generalized distributions is similar to the fundamental ($n = 1$) Gamma distribution [cf.\ Fig.~\ref{fig:setup}(b)], but the tails become even heavier with increasing order $n$, for the same mean value, reflecting the non-linear nature of the multiphoton emission process. Furthermore, the BSV light is of multimode nature in our case, which directly impacts the photon number distribution~\cite{Allevi2015}: To obtain the theoretical $m$-th order multi-mode distribution, we have to convolve $m$ single-mode probability distributions because they are statistically independent.

The influence of both the non-linearity and the number of modes on the obtained distributions is shown in Fig.~\ref{fig:BSV}(b-d). We first investigate the case with many spatial modes and a narrow spectrum [Fig.~\ref{fig:BSV}(b)] for the highest BSV pulse energy of $E_p= 18$\,nJ and an experimental mean value of $\mu = 2.6$. We compute the convolution integrals of the Gamma functions with the non-linearity of $n=4$ fixed and vary the number of modes $m$ to infer which $m$ values reproduce the experimental curves the best. We find the best agreement for $m=11\pm3$ (orange), clearly deviating from the single-mode distribution (red).

Now we close the slit down to a few hundred micrometer, which reduces the number of spatial modes but also the available pulse energy ($E_p = 6$\,nJ maximum), see Fig.~\ref{fig:BSV}(c). Here we also plot the analytical distributions for $m=1$ (red) and $m=2$ (orange), still with $n=4$. Clearly, the experimental distribution is now very close to the single-mode fourth-order distribution. Notably, we observe large electron number fluctuations reaching up to extreme events of 65 electrons per pulse, despite the small mean value of $\mu = 0.27$. We note that the probability for detecting such an event with a Poissonian distribution with a mean of $0.27$ equals $10^{-128}$.

The number of time/frequency modes is given by the ratio between the detector resolution and the coherence time of the light, in our case the BSV \cite{mandel1995optical,Ivanova2006}. Therefore, a single-mode distribution can only be observed if the response time of the detector, in the case here the tip, is smaller than the coherence time of the light. The fact that we observe the single-mode Gamma distribution for the electrons indicates that the electron emission follows the ultrafast fluctuations present in broadband BSV light. This is because the electron emission process is prompt on the (sub-) femtosecond time scale~\cite{Hommelhoff2006,Piglosiewicz2013,Dienstbier2023}, which is small compared to the coherence time of $\sim 170$\,fs of the BSV light, corresponding to the spectral bandwidth of 50\,nm. We note that, similar to averaging over many pulses, different temporal emission events within one pulse do not lead to multiple modes as long as intensity fluctuations of the light are not washed out. Clearly, such a (sub-) femtosecond response time is out of reach for even the fastest photodiodes. For comparison, such a photodiode could only detect the averaged photon number distribution of roughly $\sim 20$ temporal modes present for the given bandwidth~\cite{Spasibko2017}. 

Interestingly, already small admixtures of a few percent Poissonian-governed emission would change the single-mode electron number distribution and shift the maximum of the distribution to larger $N_e$ values (see Methods). While field-emitted electrons show Poissonian emission statistics~\cite{kiesel2002}, our measurements indicate that the quantum statistics of the light field alone determines the electron number distribution for (multi-photon) photoemission, similar to optical harmonics generated from BSV light~\cite{Manceau2019}.

To contrast the single-mode scenario, we show in Fig.~\ref{fig:BSV}(d) the electron number distribution obtained with neither spatially nor spectrally filtered broadband BSV light. Overall the higher pulse energy of $E_p=36$\,nJ in this case allows us to reach a higher mean value of $\mu = 48$ and up to 300 electrons per pulse. The highest probability to detect electrons is clearly shifted now to $N_e= 22$.  For this unfiltered configuration, we find the best agreement between experimental and multi-mode gamma distribution for $m = 57\pm5$ (orange curve).

Clearly, the extreme events compared to the mean value are heavily reduced here as the distribution gradually evolves into a Poisson distribution. This decrease in fluctuations can be explained by the increase in the number of temporal modes: The increased bandwidth reduces the coherence time down to $\sim 14$\,fs, which is still much larger than the emission time of the electrons. Yet, the spectrum of such spectrally broadband BSV light (inset in Fig.~\ref{fig:BSV}(d)) adds another degree of complexity. In this case, group delay dispersion (GDD) cannot be neglected anymore. Due to optical elements along the optics beam path, a GDD of roughly $-400\pm100~\mathrm{fs^2}$ at the central wavelength accumulates. For a coherent Fourier-limited laser pulses with the same bandwidth the initial pulse duration (13\,fs) would broaden by a factor of $\sim 5-6$, which would not change the Poissonian electron number distribution. For the BSV case, however, different temporal modes disperse and mix at the tip apex. In other words, initially correlated photon pairs do not arrive at the same time anymore. Hence, the tip, being a prompt detector with a sub-femtosecond response time as shown above, tells these events apart, which is why they do not lead to a spike in the multi-photon emission as they would if the photons arrived  simultaneously, i.e., within the tip's response time. As an effect, the tip detector averages over these dispersed modes~\cite{Kopylov2020}, which explains the increase in modes from $m=11$ (b) to $m=57$ (d). The ratio of these mode numbers roughly equaling 5 is of similar size as the temporal broadening we would expect for coherent light pulses, governed by dispersive effects.

In summary, our results show that we can imprint various photon number distributions on photo-emitted electrons. With coherent femtosecond laser pulses, we observe clean Poissonian electron number statistics. For the non-classical BSV light, we can vary the counting statistics of emitted electrons depending on the number of spatial and temporal modes of the BSV present at the tip apex. Most importantly, we have shown that the emitted electrons inherit the driving light's photon number statistics. We expect this to hold not only for super-Poissonian statistics like shown here but also for sub-Poissonian statistics, which may have direct implications for electron imaging applications: Both in light and electron optics, the imaging quality directly benefits from this, as the Fano factor, i.e., the ratio between the variance and the mean, becomes smaller than 1 for sub-Poissonian statistics. This might find direct applications in advanced electron microscopy.

We believe that this is the first time that non-classical electron statistics are generated through the electron emission process and not via filtering~\cite{Keramati2021,haindl2022coulomb,meier2022few}, which almost always comes at the cost of loosing brightness. More generally, our work paves the way to tailor-make photon distributions and transfer them into the electron realm (and vice versa \cite{Pizzi2023}), not only to venture deeper into quantum electron optics, but also for various applications beyond shot noise-reduced imaging. Last, the intensities present in our experiment are large enough to drive electrons strongly in the light field, opening the door to the nascent field of strong-field electron quantum optics \cite{EvenTzur2023}.

\subsection*{Acknowledgements}
This research was supported by the European Research Council (Consolidator Grant NearFieldAtto and Advanced Grant AccelOnChip) and the Deutsche Forschungsgemeinschaft (DFG, German Research Foundation) – Project-ID 429529648 – TRR 306 QuCoLiMa ("Quantum Cooperativity of Light and Matter") and Sonderforschungsbereich 953 ("Synthetic Carbon Allotropes"), Project-ID 182849149. J.H.\ acknowledges funding from the Max Planck School of Photonics.
\FloatBarrier
\newpage

\newpage
\FloatBarrier
\section*{Methods}\label{secA1}
\setcounter{figure}{0} 
\renewcommand{\figurename}{Extended Data Fig.}

\subsection*{BSV light generation}
We here give more details on the generation of the BSV light. To generate BSV, we focus the pump (pulse duration 1.6\,ps, energy per pulse up to 0.4\,mJ) with a 700-mm cylindrical lens into the 10\,mm beta-barium borate (BBO) crystal cut for collinear degenerate type-I phase matching. The beam remains broad ($\sim$3\,mm) in the optic axis plane, which eliminates the spatial walk-off effect. The orientation of the crystal is set for a nearly collinear degenerate regime. We note that in the same configuration various other unwanted accompanying processes inside the same crystal, such as PDC saturable absorption, second harmonic and sum frequency generation could be observed \cite{Spasibko2020}. Such effects might result in an uncontrolled change of BSV statistics or even the intensity reduction. To alleviate the impact of these processes we slightly detune the crystal from its optimal PDC phase-matching angle reducing on purposely the BSV intensity and keep the maximum average pump power below 2\,W. The pump laser power is kept fixed in all the measurements.  To suppress the pump beam after the crystal we use a dichroic mirror and bandpass filters with a total optical density of 22 at 800\,nm.

\subsection*{Effect of Poissonian influence on single-mode Gamma distributions}
\begin{figure}[h]
    \centering
    \includegraphics[width = 0.8\linewidth]{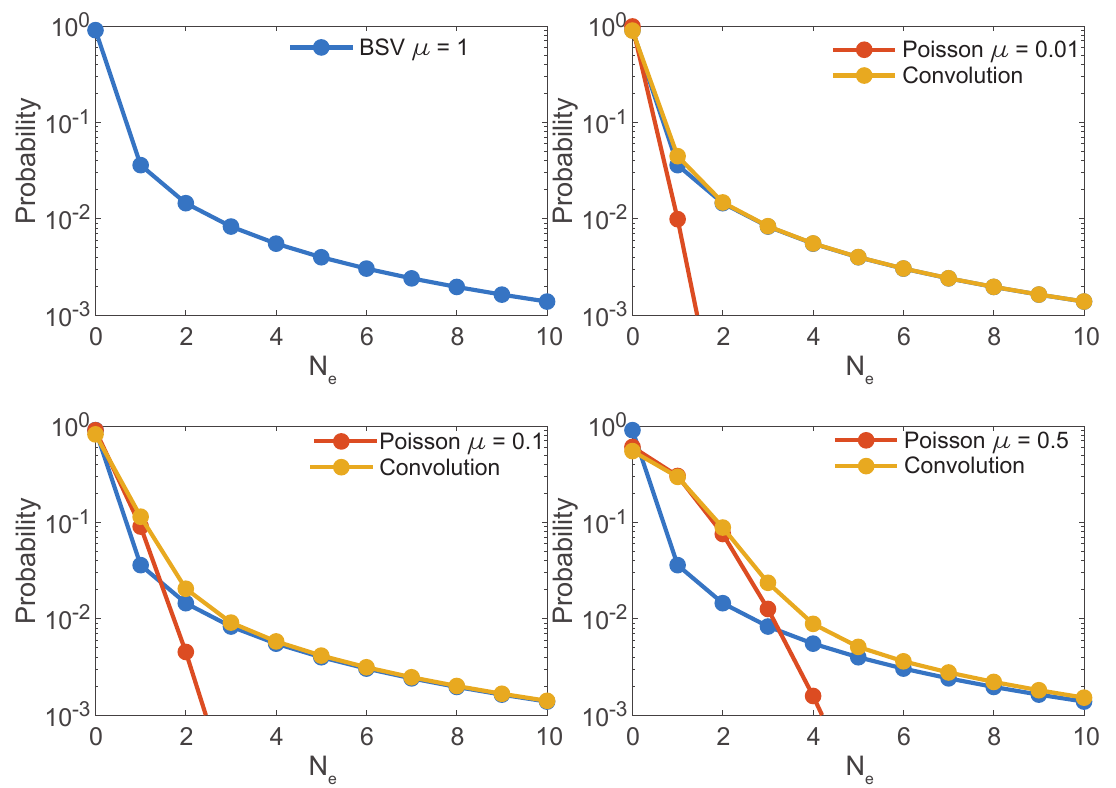}
    \caption{(a) Electron number distribution for single-mode fourth-order Gamma distribution (blue) for an average of $\mu = 1$ electrons per laser pulse, calculated from Eq.~(\ref{eq:gamma_2}) (see main text). Lines are guide to the eye. (b)-(d) Blue are the distributions from (a). Orange are the Poisson distributions corresponding to a mean with (b) $\mu = 0.01$, (c) $\mu = 0.1$ and (d) $\mu = 0.5$. The yellow curves are the corresponding convolutions of the PDC and the Poisson distribution. For increasing Poissonian admixture we observe a clear deviation from the original distribution in (a). \label{fig:admixture}}
\end{figure}

As we have shown in the main text, we were able to imprint the statistics of BSV light onto the electron number distribution. The measured fourth-order single-mode distribution of the electrons requires that the electron emission is mainly steered by the light field, which we show in the following. In Extended Data Fig.~\ref{fig:admixture}(a) we show the theoretical single-mode fourth-order Gamma  distribution (see Eq.~(\ref{eq:gamma_2})) of electrons with a mean value $\mu = 1$. We now assume that the electrons have an additional uncorrelated contribution like in the DC field-emitted case, showing a Poissonian electron number distribution. When we mix the single-mode distribution and the Poissonian distribution by convoluting the both, we obtain the yellow curve in Extended Data Fig.~\ref{fig:admixture}(b-d). For 1\% (b) admixture of Poissonian contribution, only small differences to the original distribution are observed. Deviations from the single-mode case start already at 10\% admixture, visible for electron events up to $N_e = 3$. For 50\% admixture the resulting distribution clearly deviates from the single-mode case.

Furthermore, the single-mode distribution in the experiment requires that the electron bath at the tip apex acts collectively (coherently). If the tip apex acted as $k$ independent emitters, we would get the incoherent sum of these $k$ emission sites, being again the $k$-times convolution of the single-mode distribution.

\FloatBarrier
\newpage

\bibliography{literature}

\end{document}